\documentclass[aps,prd,
amsmath,amssymb,nofootinbib,superscriptaddress]{revtex4-2}
\usepackage{graphicx}  
\usepackage{color}
\usepackage{times}
\usepackage[utf8]{inputenc} 
\usepackage{bm}
\usepackage{ulem}
\usepackage{multirow}
\usepackage{makecell}
\usepackage{url}
\usepackage{natbib}
\usepackage{mathrsfs}
\usepackage{physics}
\usepackage{comment}
\usepackage[table,xcdraw]{xcolor}
\usepackage{booktabs,makecell}
\usepackage{amsmath}
\usepackage{pifont}
\usepackage[colorlinks=true,citecolor=blue,urlcolor=blue,linkcolor=blue]{hyperref}
\usepackage{microtype}
\usepackage[none]{hyphenat}

\newcommand{\dalm}{\kern1pt\vbox{\hrule height 0.9pt\hbox{\vrule width 0.9pt
\hskip 2.5pt\vbox{\vskip 5.5pt}\hskip 3pt\vrule width 0.3pt}\hrule height 0.3pt}
\kern1pt}

\begin{document}



\title{Universal relations between the quasinormal modes of neutron stars and magnetic tidal deformability}

\author{Hajime Sotani}
\email{sotani@yukawa.kyoto-u.ac.jp}
\affiliation{Department of Mathematics and Physics, Kochi University, Kochi, 780-8520, Japan}
\affiliation{RIKEN Center for Interdisciplinary Theoretical and Mathematical Sciences (iTHEMS), RIKEN, Wako 351-0198, Japan}
\affiliation{Theoretical Astrophysics, IAAT, University of T\"{u}bingen, 72076 T\"{u}bingen, Germany}


\date{\today}

\begin{abstract}
Tidal deformabilities are one of the observable quantities characterizing neutron stars, which are strongly associated with the stellar compactness, the ratio of the stellar mass to the radius. In addition to the tidal deformability, the quasinormal modes excited in a neutron star are also an important property for extracting information about the neutron star interior, adopting gravitational wave asteroseismology. In this study, we especially focus on the magnetic tidal deformability, which acts on the gravitational waveform from a neutron star binary merger as a higher-order effect than the electric tidal deformability, and derive the universal relations expressing the quasinormal modes, such as the fundamental ($f$-), 1st pressure ($p_1$-), and 1st spacetime ($w_1$-) modes, as a function of the magnetic tidal deformability. The universal relations derived in this study exhibit accuracy more or less comparable to those of the electric tidal deformability. 
\end{abstract}

\maketitle


\section{Introduction}
\label{sec:I}

Neutron stars formed via supernova explosions are unique objects for probing physics under extreme states. The density inside the star easily exceeds the nuclear saturation density, and its gravitational and magnetic fields become much stronger than those observed in our solar system~\cite{ST83}. Since such extreme conditions are quite difficult to realize on Earth, one might inversely extract the imprint of physics under such extreme conditions through the observations of neutron stars. For instance, the discoveries of $2M_\odot$ neutron stars exclude the soft equation of state (EOS), with which the expected maximum mass cannot reach the observed masses~\cite{D10,A13,C20,F21}. In particular, the mass of black widow, PSR J0952-0607, is estimated with $M = 2.35 \pm 0.17 M_\odot$~\cite{Romani22}, which is the heaviest neutron star known in the Milky Way. The discovery of a higher mass of a neutron star gives us more severe constraints on the EOS. 

The gravitational waves observed during the neutron star merger, GW170817, also enabled us to constrain the dimensionless (electric) tidal deformability~\cite{GW170817}, which leads to a constraint on a $1.4M_\odot$ neutron star radius, i.e., $R_{1.4}\le 13.6$ km~\cite{Annala18}. Now, gravitational waves have become a new tool for obtaining astronomical information, alongside electromagnetic waves and neutrinos. Moreover, since the photon from a neutron star's surface bends due to the strong gravitational field induced by the neutron star, which is one of the relativistic effects, one may primarily constrain the stellar compactness, $M/R$, with the stellar mass $M$ and radius $R$~\cite{PFC83,LL95,PG03,PO14,SM18,Sotani20a}. In practice, the mass and radius for PSR J0030+0451~\cite{Riley19,Miller19} and PSR J0740+6620~\cite{Riley21,Miller21} are estimated via the observations by the Neutron star Interior Composition Explorer (NICER) operating on the International Space Station (ISS), and the constraint on PSR J0030+0451 has been further updated~\cite{Vinciguerra24}. Meanwhile, the quasi-periodic oscillations observed in the magnetar flares are considered to be associated with neutron star oscillations. By identifying the observed frequencies with the crustal torsional oscillations, with the help of the constraints on the nuclear parameters obtained from the terrestrial experiments, the mass and radius of GRB 200415A are estimated~\cite{SKS23}. These neutron star observations generally give us constraints on the EOS for a relatively higher density region, while the terrestrial nuclear experiments also make constraints on the EOS for a lower density region~\cite{SNN22,SO22,SN23}. Developments in technology improve the precision of astronomical observations and ground-based experiments, which will impose stricter constraints on the EOS for neutron star matter.

In addition to the stellar (static) properties, such as mass and radius, the oscillation frequencies from neutron stars provide important information for extracting properties in extreme states. Since the oscillation frequency strongly depends on the stellar properties, one could obtain the signature of the interior information via the observations of their frequencies. This technique is known as asteroseismology (or gravitational wave asteroseismology), which is similar to seismology on Earth and helioseismology on the Sun. In fact, various oscillation modes can be excited in a neutron star, depending on the input physics. Therefore, by identifying the observed frequency with a specific mode, one could extract the imprint of the corresponding physics. For example, since the fundamental ($f$-) modes are acoustic oscillations, those frequencies are strongly associated with the average density of the star. That is, the observation of the $f$-mode frequencies gives us the information of the stellar average density~\cite{AK1996,AK1998}. Actually, the quasi-periodic oscillations observed in the magnetars, considered as a result of the neutron star oscillations, could be identified with the neutron star crustal torsional oscillations, which gives us constraints on the neutron star crustal EOS and/or the stellar models~\cite{SKS23,GNHL2011,SNIO2012,SIO2016,Sotani24a}. Once one observes the frequencies of the gravitational waves from a neutron star, one could constrain the neutron star mass, radius, EOS, and also rotational properties, e.g., Refs.~\cite{STM2001,SH2003,TL2005,SYMT2011,PA2012,DGKK2013,Sotani20b,Sotani21,KHA15,SD22,SD24}. Furthermore, this technique may also enable us to understand the supernova gravitational waves, e.g., Refs.~\cite{FMP2003,FKAO2015,ST2016,ST2020a,SKTK2017,MRBV2018,SKTK2019,TCPOF19,SS2019,ST2020,STT2021,SMT24}.

The observations of neutron star properties and/or the oscillation frequencies are definitely important for revealing neutron star physics, but, in general, one cannot avoid the dependence of the EOS for discussing such properties. Nevertheless, it is also known that there are relations between some specific combinations of neutron star properties, including its oscillation frequencies, which are almost independent of the EOS. These relations are known as universal relations (or empirical relations). For instance, the relations between the momentum inertia, $I$, the tidal Love numbers, and the spin-induced quadrupole moment, $Q$, hardly depend on the EOS for neutron star matter \cite{Maselli13,Yagi14,GGRB21}. Similarly, the universal relations with neutron star oscillation frequencies are also found as in Refs. \cite{AK1996,AK1998,TL2005,Chan14,SK21,Sotani22,Zhao22,PVC23}. These universal relations, if any, are quite important to estimate the neutron star properties even though the true EOS remains undetermined, and also to validate the theory of gravity. Especially, the relation with the $f$-mode frequencies may be more important for considering gravitational waves from neutron star binary mergers, because it is crucial to understand the waveforms, including the resonance with the orbital frequency just before the merger~\cite{SKK24}. Meanwhile, the gravitational waves from the neutron star merger also depend not only on the electric tidal deformation but also, albeit to an extremely small extent, on the magnetic one. Thus, this study investigates the universality between the magnetic tidal deformability and the quasinormal modes excited in neutron stars. 

This manuscript is organized as follows. In Sec. \ref{sec:MTD}, we briefly describe the tidal deformability. In Sec. \ref{sec:Universal}, we discuss the relation between quasinormal modes of neutron stars and magnetic tidal deformability. Finally, we summarize our findings in this study and the conclusions in Sec. \ref{sec:Conclusion}. Unless otherwise mentioned, we adopt geometric units with $c=G=1$, where $c$ and $G$ denote the speed of light and the gravitational constant, and use the metric signature $(-,+,+,+)$.

\section{Magnetic tidal deformability}
\label{sec:MTD}

A neutron star in a close orbit system can tidally deform due to the gravitational force from the companion, where the tidal field is decomposed into electric, ${\cal E}_L$, and magnetic components, ${\cal M}_L$. These tidal fields are associated with the mass multipole moment ${\cal Q}_L$ and a current multipole moment, ${\cal S}_L$, as
\begin{gather}
  {\cal Q}_L=\lambda_\ell {\cal E}_L, \\
  {\cal S}_L=\sigma_\ell {\cal M}_L,   
\end{gather}
where $\lambda_\ell$ and $\sigma_\ell$ are the $\ell$-th gravitoelectric and gravitomagnetic
tidal deformabilities, respectively \cite{DN09,PC21}. Using $\lambda_\ell$ and $\sigma_\ell$, the dimensionless gravitoelectric and gravitomagnetic tidal Love numbers, $k_\ell$ and $j_\ell$, are given by \cite{PC21}
\begin{gather}
  k_\ell=\frac{(2\ell-1)!!}{2} \frac{\lambda_\ell}{R^{2\ell+1}}, \\
  j_\ell=4(2\ell-1)!! \frac{\sigma_\ell}{R^{2\ell+1}},  
\end{gather}
while the corresponding dimensionless tidal deformability is given by \cite{PC21}
\begin{gather}
  \Lambda_\ell=\frac{2k_\ell}{(2\ell-1)!!} \left(\frac{R}{M}\right)^{2\ell+1}, \\
  \Sigma_\ell=\frac{j_\ell}{4(2\ell-1)!!} \left(\frac{R}{M}\right)^{2\ell+1}.  
\end{gather}
In this study, we especially focus on the $\ell=2$ dimensionless magnetic tidal deformability, $\Sigma_2$.

Without deformation, we assume that the neutron stars are spherically symmetric, whose metric is given by
\begin{gather}
  ds^2 = -e^{2\Phi}dt^2 + e^{2\Lambda}dr^2 + r^2 \left(d\theta^2 + \sin^2 \theta d\phi^2\right),
\end{gather}
where the metric functions, $\Phi$ and $\Lambda$, depend only on the radial coordinate, $r$, and $\Lambda$ is directly connected to the mass function, $m$, via $e^{-2\Lambda}=1-2m/r$. On this background model, the non-zero components of the metric perturbations associated with the gravitomagnetic tidal deformability are given by 
\begin{gather}
  \delta g_{t\theta} = -\frac{h_0}{\sin\theta}\partial_\phi Y_{\ell m}, \ \ 
  \delta g_{t\phi} = h_0\sin\theta\partial_\theta Y_{\ell m}, \\
  \delta g_{r\theta} = -\frac{h_1}{\sin\theta}\partial_\phi Y_{\ell m}, \ \ 
  \delta g_{r\phi} = h_1\sin\theta\partial_\theta Y_{\ell m},
\end{gather}
which are the axial type perturbations and do not involve the density variation. From the linearized Einstein equation, one can derive the perturbation equation for $h_0$ as
\begin{gather}
  h''_0 - 4\pi r (p+\varepsilon)e^{2\Lambda} h'_0 - \left[\frac{\ell(\ell+1)}{r^2} - \frac{4m}{r^3} +8\pi \Theta(p+\varepsilon) \right]e^{2\Lambda}h_0 = 0, \label{eq:Mperturb} \\
  \dot{h}_0 = h'_1 + h_1(\Lambda - \Phi),
\end{gather}
where the dot and prime denote the partial derivative with $t$ and $r$, while $p$ and $\varepsilon$ are the pressure and energy density determined by solving the Tolman-Oppenheimer-Volkoff equation. In Eq.~(\ref{eq:Mperturb}), $\Theta=+1$ corresponds to the strictly static case, while $\Theta=-1$ is for irrotational fluid. As discussed in Ref.~\cite{Pani18}, since the situation with $\Theta=-1$, where the irrotational fluid is induced by tidal force, could be more realistic, we focus only on this case in this study. From the regularity condition, one can obtain the boundary condition at the stellar center as
\begin{equation}
  h_0=\frac{r}{\ell+1}h'_0.
\end{equation}

By integrating Eq.~(\ref{eq:Mperturb}) for $\ell=2$ from the stellar center to surface, the values of $h_0$ and $h_0'$ at the surface are determined. Using the resultant values of $h_0(R)$ and $h_0'(R)$, the $\ell=2$ gravitomagnetic Love number, $j_2$, is determined by 
\begin{equation}
  j_2 = \frac{24{\cal C}^5}{5\tilde{D}_2}\left[2\left(\tilde{y}_2-2\right){\cal C}-\tilde{y}_2+3\right],
\end{equation}
where ${\cal C}$ is the stellar compactness, ${\cal C}=M/R$, while $\tilde{y}_2$ and $\tilde{D}_2$ are given by
\begin{gather}
  \tilde{y}_2 = \frac{Rh'_0(R)}{h_0(R)}, \\
  \tilde{D}_2= 2{\cal C}\left[2\left(\tilde{y}_2+1\right){\cal C}^3 + 2\tilde{y}_2{\cal C}^2 + 3\left(\tilde{y}_2-1\right){\cal C} - 3\left(\tilde{y}_2-3\right)\right] + 3\left[2\left(\tilde{y}_2-2\right){\cal C} - \tilde{y}_2 + 3\right]\ln(1-2{\cal C}).
\end{gather}

To see the EOS dependence, in this study we adopt various EOS listed in Table~\ref{tab:EOS}, which are the same set of EOS adopted in Ref.~\cite{Sotani21,SK21}. DD2~\cite{DD2}, Miyatsu~\cite{Miyatsu}, and Shen~\cite{Shen} are the EOS based on the relativistic mean field approximation; FPS~\cite{FPS}, SKa~\cite{SKa}, SLy4~\cite{SLy4}, and SLy9~\cite{SLy9} are the EOS using the Skyrm-type effective interaction; and Togashi~\cite{Togashi17} is the EOS constructed with the variational method. In Table~\ref{tab:EOS}, we list the nuclear saturation parameters, $K_0$ and $L$, for each EOS; $\eta$ defined as $\eta=(K_0L^2)^{1/3}$, which is a suitable parameter for describing a low-mass neutron star~\cite{SIOO14}; the maximum mass of a static spherically symmetric neutron star, $M_{\rm max}/M_\odot$; and the $\ell=2$ dimensionless electric tidal deformability for the $1.4M_\odot$ neutron star, $\Lambda_{1.4}$. We note that, the fiducial values of $K_0$ and $L$ obtained from the terrestrial experiments are $K_0=240\pm 20$ MeV~\cite{Shlomo06,Garg18} and  $L\simeq 60 \pm 20$ MeV~\cite{Vinas14,Li19}, while $\Lambda_{1.4}$ is constrained to be $\Lambda_{1.4}=190^{+390}_{-120}$  through the observation of GW170817~\cite{GW170817}. 

\begin{table}
\caption{Nuclear parameters for the EOS adopted in this study, $K_0$, $L$, and $\eta$. The maximum mass, $M_{\rm max}/M_\odot$, for the neutron star and the $\ell=2$ dimensionless (electric) tidal deformability, $\Lambda_{1.4}$, for the $1.4M_\odot$ neutron star constructed with each EOS are also listed.} 
\label{tab:EOS}
\begin {center}
\begin{tabular}{cccccc}
\hline\hline
EOS & $K_0$ (MeV) & $L$ (MeV) & $\eta$ (MeV) & $M_{\rm max}/M_\odot$ & $\Lambda_{1.4}$  \\
\hline
DD2
 & 243 & 55.0  & 90.2  & 2.41 & 774.8 \\ 
Miyatsu
 & 274 &  77.1 & 118  & 1.95 & 601.0 \\
Shen
 & 281 & 111  & 151  &  2.17 & 1104.0 \\  
FPS
 & 261 & 34.9 & 68.2  & 1.80 & 182.0 \\  
SKa
 & 263 & 74.6 & 114 & 2.22 & 618.0 \\ 
SLy4
 & 230 & 45.9 &  78.5 & 2.05 & 321.7 \\ 
SLy9
 & 230 & 54.9 &  88.4 & 2.16 & 469.3 \\  
Togashi
 & 245  & 38.7  & 71.6 & 2.21 & 309.2  \\ 
\hline \hline
\end{tabular}
\end {center}
\end{table}

\begin{figure}[tbp]
\begin{center}
\includegraphics[scale=0.6]{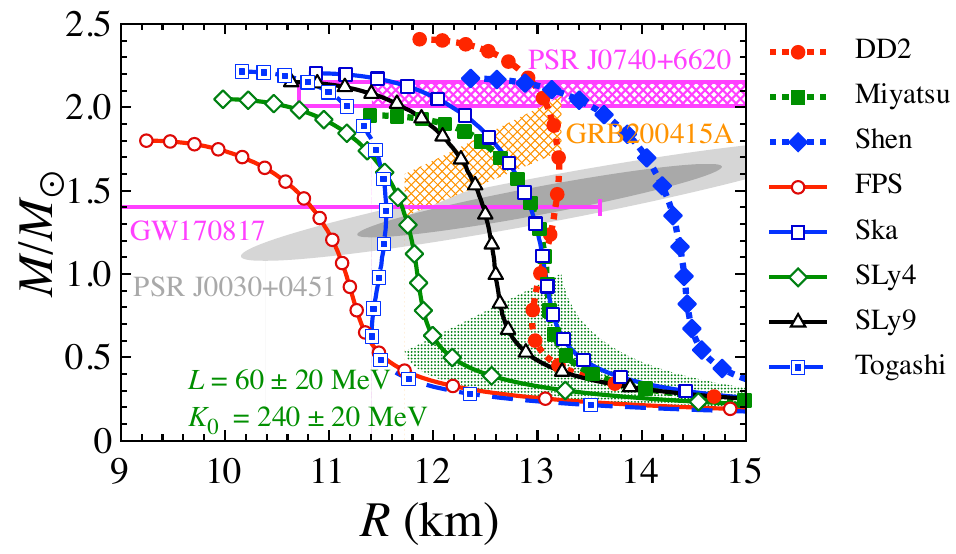} 
\end{center}
\caption{
Mass-radius relations for the neutron star models constructed with the EOS adopted in this study. For reference, we also plot the constraints obtained from the astronomical observations, i.e, PSR J0030+0451; PSR J0740+6620; GW170817; and GRB200415A, and the expected region assuming the fiducial values of nuclear saturation parameters such as $L=60\pm 20$ MeV and $K_0=240\pm 20$ MeV in the right-bottom region (see text for details). 
}
\label{fig:MR}
\end{figure}

In addition, Fig.~\ref{fig:MR} shows the relation between the mass and radius of neutron stars constructed with the EOS adopted in this study. For reference, we also show the constraints obtained from the astronomical observations. The direct observation of gravitational waves from the neutron star merger, GW170817, gives us a constraint on the dimensionless electric tidal deformability, which leads to the radius constraint of a $1.4M_\odot$ neutron star, i.e., $R_{1.4}\le 13.6$ km~\cite{GW170817,Annala18}. The pulse profile of the rotating neutron stars could tell us the stellar compactness, $M/R$, because the photon radiating from the stellar surface would bend due to the strong gravitational field induced by the neutron star as a relativistic effect. Actually, through the careful observations of pulse profiles by NICER, the mass and radius of PSR J0030+0451~\cite{Riley19,Miller19} and PSR J0740+6620~\cite{Riley21,Miller21} are constrained. The mass and radius of GRB 200415A are also evaluated by identifying the observed magnetar quasi-periodic oscillations with the crustal torsional oscillations~\cite{SKS23}. Together with the constraint from the astronomical observations, the lower-density region of a neutron star may be possible to constrain from the terrestrial experiments. Once one adopts the fiducial values of $K_0$ and $L$ mentioned before, the mass and radius are expected in the right-bottom region in the figure, using the low-mass neutron star mass formula derived in Ref.~\cite{SIOO14}. Considering these constraints with astronomical observations and fiducial values of nuclear saturation parameters, some of the EOS adopted here have already been excluded, but we consider the EOS listed in Table~\ref{tab:EOS} in this study to examine the EOS dependence in a wider parameter space.

\begin{figure}[tbp]
\begin{center}
\includegraphics[scale=0.6]{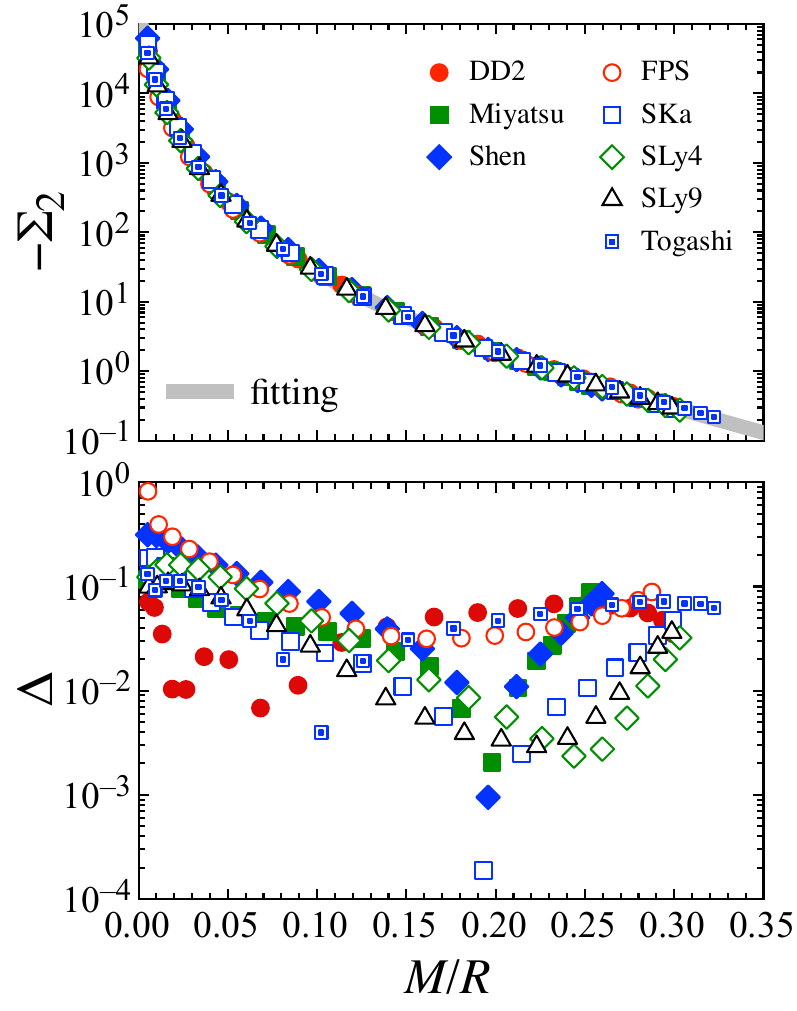} 
\end{center}
\caption{
$-\Sigma_2$ is shown as a function of the stellar compactness, $M/R$, for various EOS, where the thick solid line denotes the fitting formula given by Eq.~(\ref{eq:Sigma_MR}). The bottom panel shows the absolute values of the relative deviation of $-\Sigma_2$ from the fitting formula.
}
\label{fig:Sigma_MR}
\end{figure}

\begin{figure}[tbp]
\begin{center}
\includegraphics[scale=0.6]{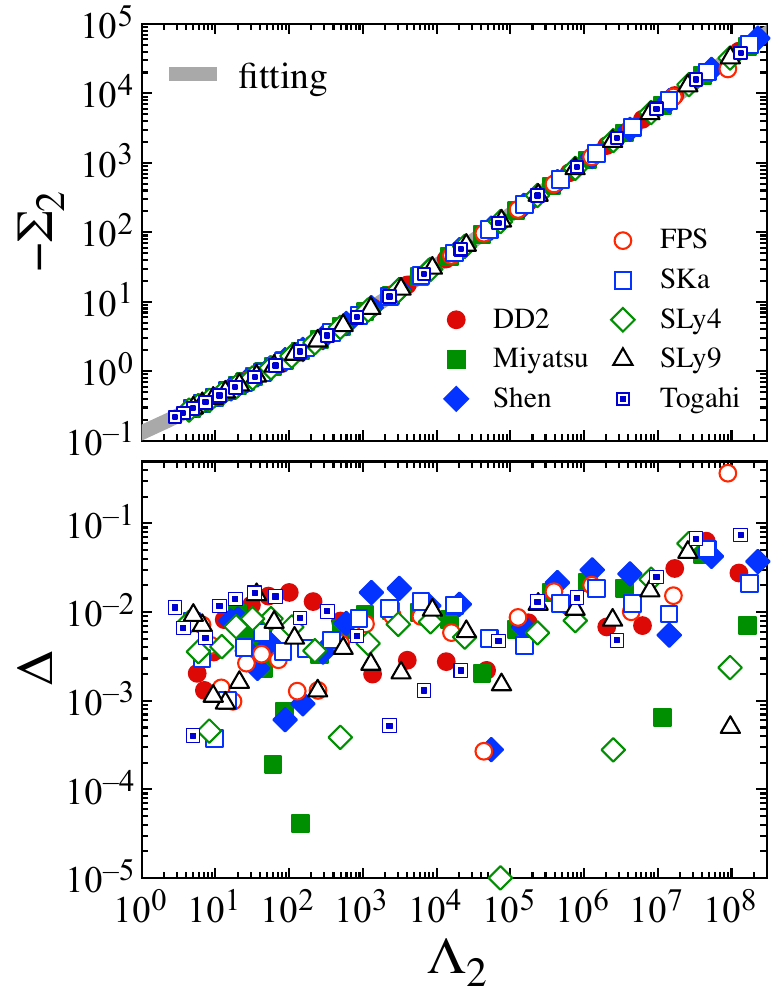} 
\end{center}
\caption{
$-\Sigma_2$ is shown as a function of $\Lambda_2$ for various EOS, where the thick solid line denotes the fitting formula given by Eq.~(\ref{eq:Sigma_Lambda}). In the bottom panel, the relative deviation from the fitting formula is also shown. 
}
\label{fig:Sigma_Lambda}
\end{figure}

It is well known that the $\ell=2$ dimensionless electric tidal deformability, $\Lambda_2$, is expressed as a function of stellar compactness~\cite{Maselli13,PVC23}. In a similar way, the $\ell=2$ dimensionless magnetic tidal deformability, $\Sigma_2$, can be expressed as a function of stellar compactness as
\begin{equation}
   \log_{10}\left(-\Sigma_2\right) = -0.002180/\tilde{\cal C}+6.0235-8.1928\tilde{\cal C}^{1/2}+3.0619\tilde{\cal C}-0.3505\tilde{\cal C}^2, \label{eq:Sigma_MR}
\end{equation}
where $\tilde{\cal C}={\cal C}/0.172$. Here, the constant used for normalization, 0.172, is the compactness for the neutron star model with $1.4M_\odot$ and 12 km. As shown in Fig.~\ref{fig:Sigma_MR}, using this fitting formula, one can expect the absolute value of $\Sigma_2$ within $\sim 10 \%$  for a canonical neutron star model, whose compactness is larger than $\sim 0.17$. Owing to the universality of $\Lambda_2$ and $\Sigma_2$ with respect to the stellar compactness, one can easily expect the existence of the universality between $\Lambda_2$ and $\Sigma_2$~\cite{Yagi14,PVC23}. In fact, such a relation is given by
\begin{equation}
   \log_{10}\left(-\Sigma_2\right)=-0.8843 + 0.4808\chi + 0.028172 \chi^2 +0.0014241 \chi^3-0.0002369 \chi^4, \label{eq:Sigma_Lambda}
\end{equation}
where $\chi=\log_{10}\Lambda_2$. As shown in Fig.~\ref{fig:Sigma_Lambda}, this relation tells us the value of $\Sigma_2$ within a few per cent for the canonical neutron stars, whose $\Lambda_2$ is less than 1000.

\section{Universal relation between quasinormal modes and magnetic tidal deformability}
\label{sec:Universal}

\begin{figure*}[tbp]
\begin{center}
\includegraphics[scale=0.6]{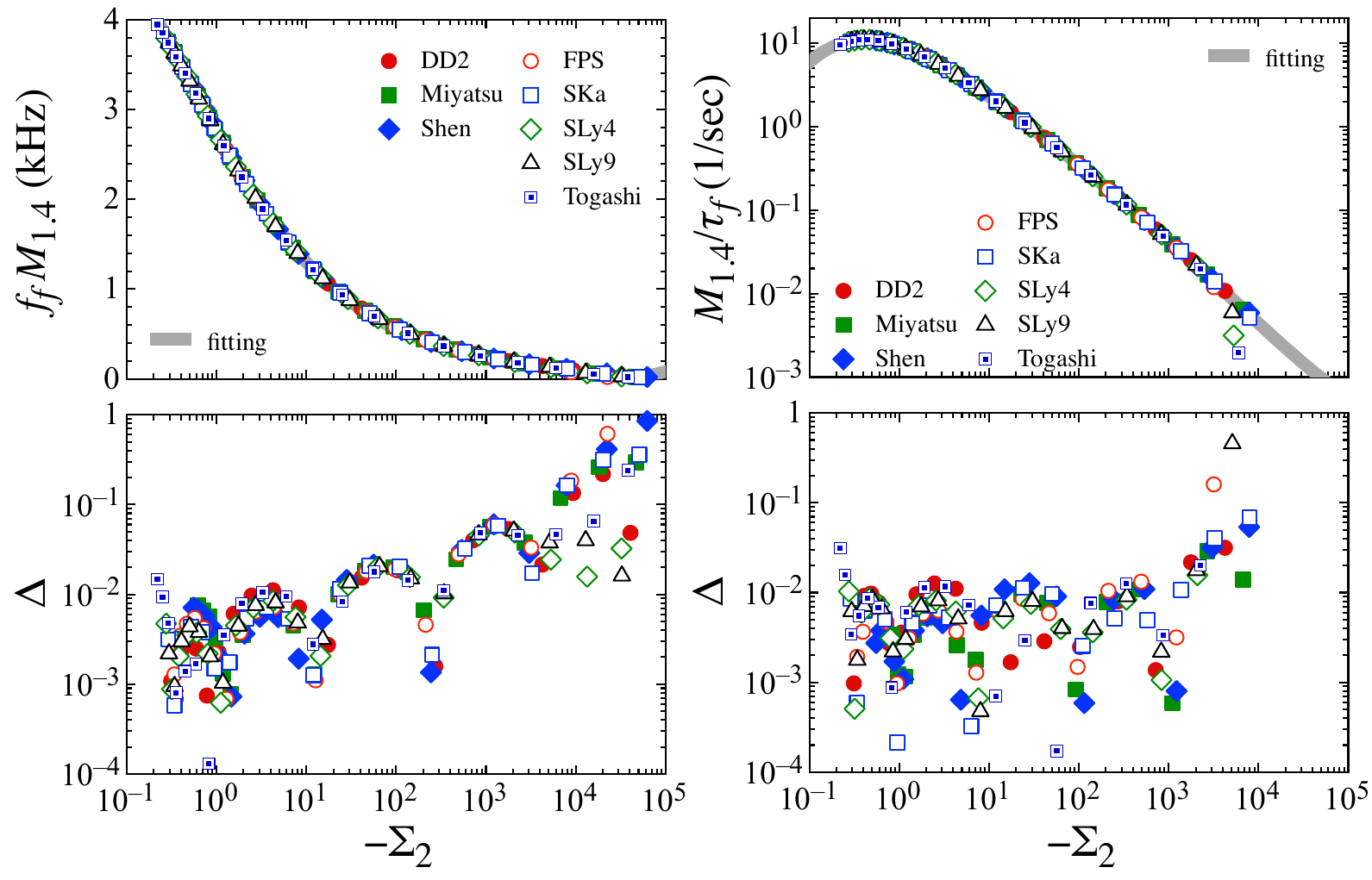} 
\end{center}
\caption{
The mass-scaled $f$-mode frequencies (top-left panel) and mass-scaled damping rate (top-right panel) are shown as a function of $-\Sigma_2$. In both panels, the solid lines denote the fitting formulae given by Eqs.~(\ref{eq:ffM-Sigma}) and (\ref{eq:ftauM-Sigma}). The bottom panels show the relative deviation from the fitting formulae, i.e., $\Delta=|A^{\rm N}-A^{\rm F}|/A^{\rm N}$ with the values determined numerically, $A^{\rm N}$, and those estimated from the fitting formulae, $A^{\rm F}$. 
}
\label{fig:ffM-ftauM}
\end{figure*}

Now, we discuss the relations between the frequencies of quasinormal modes excited in neutron stars and the $\ell=2$ magnetic tidal deformability, $\Sigma_2$. The quasinormal modes are determined by solving the eigenvalue problems with the perturbation equations derived from the linearized Einstein equations, where the matter and metric perturbations are coupled with each other, adopting appropriate boundary conditions. The details of the perturbation equations, boundary conditions, and how to numerically solve the eigenvalue problems are shown in Refs.~\cite{STM2001,ST2020}. The frequencies of the quasinormal modes generally depend on the EOS as well as the stellar properties. Nevertheless, it is shown that they can be well expressed as a function of the electric tidal deformability almost independently of the EOS~\cite{Chan14,SK21,PVC23}. In the same way, as shown in Fig.~\ref{fig:ffM-ftauM}, we find that the mass-scaled $f$-mode frequencies, $M_{1.4}f_f$ (top-left panel), and their mass-scaled damping rate, $M_{1.4}/\tau_f$ (top-right panel), can also be well expressed as a function of the magnetic tidal deformability, where $M_{1.4}\equiv M/(1.4M_\odot)$, $f_f$ denotes the $f$-mode frequencies, and $\tau_f$ denotes the damping time of the $f$-mode oscillations. In this figure, the solid lines denote the fitting formulae given by
\begin{gather}
  f_fM_{1.4}\ ({\rm kHz})
   = 2.7464 -1.7864x+ 0.2781x^2+ 0.096669x^3 -0.03856x^4+ 0.0036256x^5, \label{eq:ffM-Sigma} \\
  \log_{10}(M_{1.4}/\tau_f\ (1/{\rm sec}))
   = 0.9650 -0.3548x -0.3673x^2+ 0.1472x^3-0.031854x^4+  0.0026456x^5, \label{eq:ftauM-Sigma}
\end{gather}
where $x\equiv \log_{10}(-\Sigma_2)$. In the bottom panels of Fig.~\ref{fig:ffM-ftauM}, we show the absolute value of the relative deviation of the frequencies and damping rate from the estimations with the fitting formulae. From this figure, one can observe that the fitting formulae derived here can work for estimating the $f$-mode frequency and damping rate within a few percent accuracy for a canonical neutron star model, whose $-\Sigma_2$ is less than $\sim 10$. We note that the damping rate for extremely low-mass neutron stars significantly deviates from the fitting formula (Eq.~(\ref{eq:ftauM-Sigma})). This is because the dependence of the damping rate on the stellar models dramatically changes due to the avoided crossing between the $f$- and the 1st pressure ($p_1$-) modes for a neutron star with extremely low mass \cite{Sotani21}.

\begin{figure}[tbp]
\begin{center}
\includegraphics[scale=0.6]{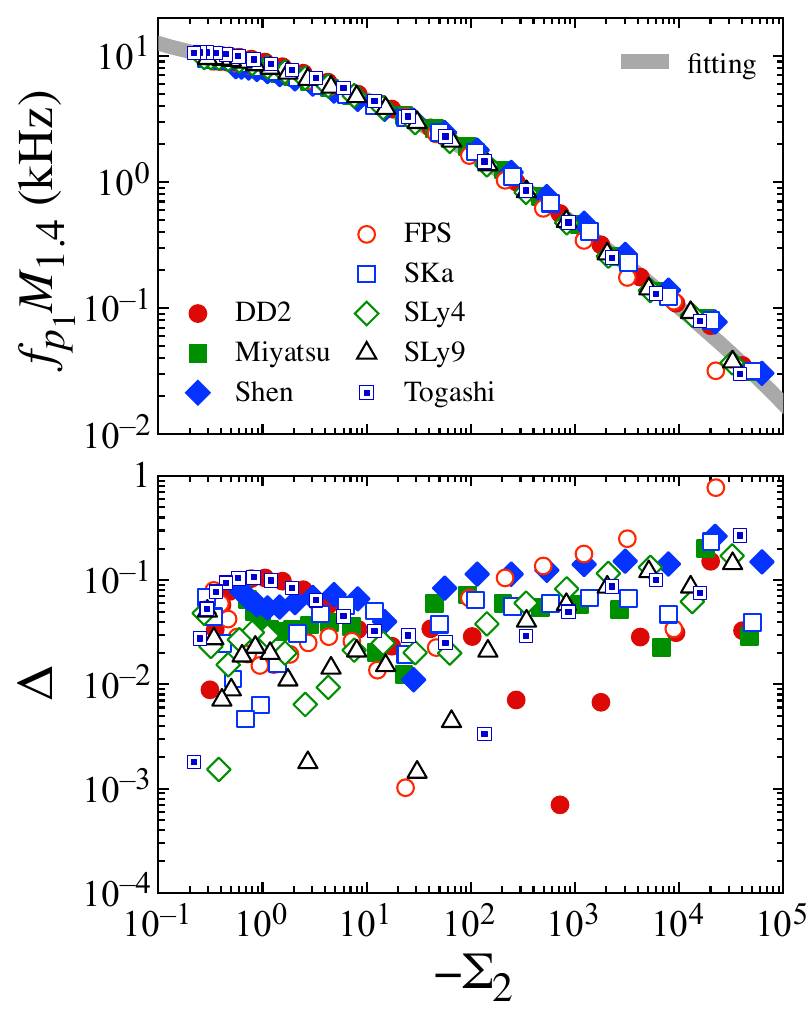} 
\end{center}
\caption{
Same as in Fig.~\ref{fig:ffM-ftauM}, but for the $p_1$-mode frequencies with the fitting formula given by Eq.~(\ref{eq:fp1M-Sigma}). 
}
\label{fig:fp1M}
\end{figure}

The behavior of the $p_1$-mode frequencies is more complicated because their eigenfunctions have one node inside the star, and the position of the node depends on the interior structure. Therefore, the fitting of the $p_1$-mode frequencies with stellar properties is more difficult than that for the $f$-modes. Nonetheless, as shown in Fig.~\ref{fig:fp1M}, the mass-scaled $p_1$-mode frequencies, $f_{p_1}M_{1.4}$, have a strong association with the magnetic tidal deformability, where the solid line denotes the fitting formula given by
\begin{equation}
  \log_{10}(f_{p_1}M_{1.4}\ ({\rm kHz}))
   = 0.9059 -0.1875x-0.039707x^2-0.034363x^3+ 0.010073x^4-0.0008749x^5. \label{eq:fp1M-Sigma}
\end{equation}
The bottom panel of Fig.~\ref{fig:fp1M} shows the absolute values of the relative deviation from the fitting formula. From this figure, we find that the mass-scaled $p_1$-mode frequencies can be estimated from the magnetic tidal deformability within $\sim 10\%$ accuracy for a canonical neutron star model. We also attempt to derive the fitting formula for the $p_1$-mode damping rate as a function of $-\Sigma_2$, but unfortunately, we were unable to find it.

\begin{figure*}[tbp]
\begin{center}
\includegraphics[scale=0.6]{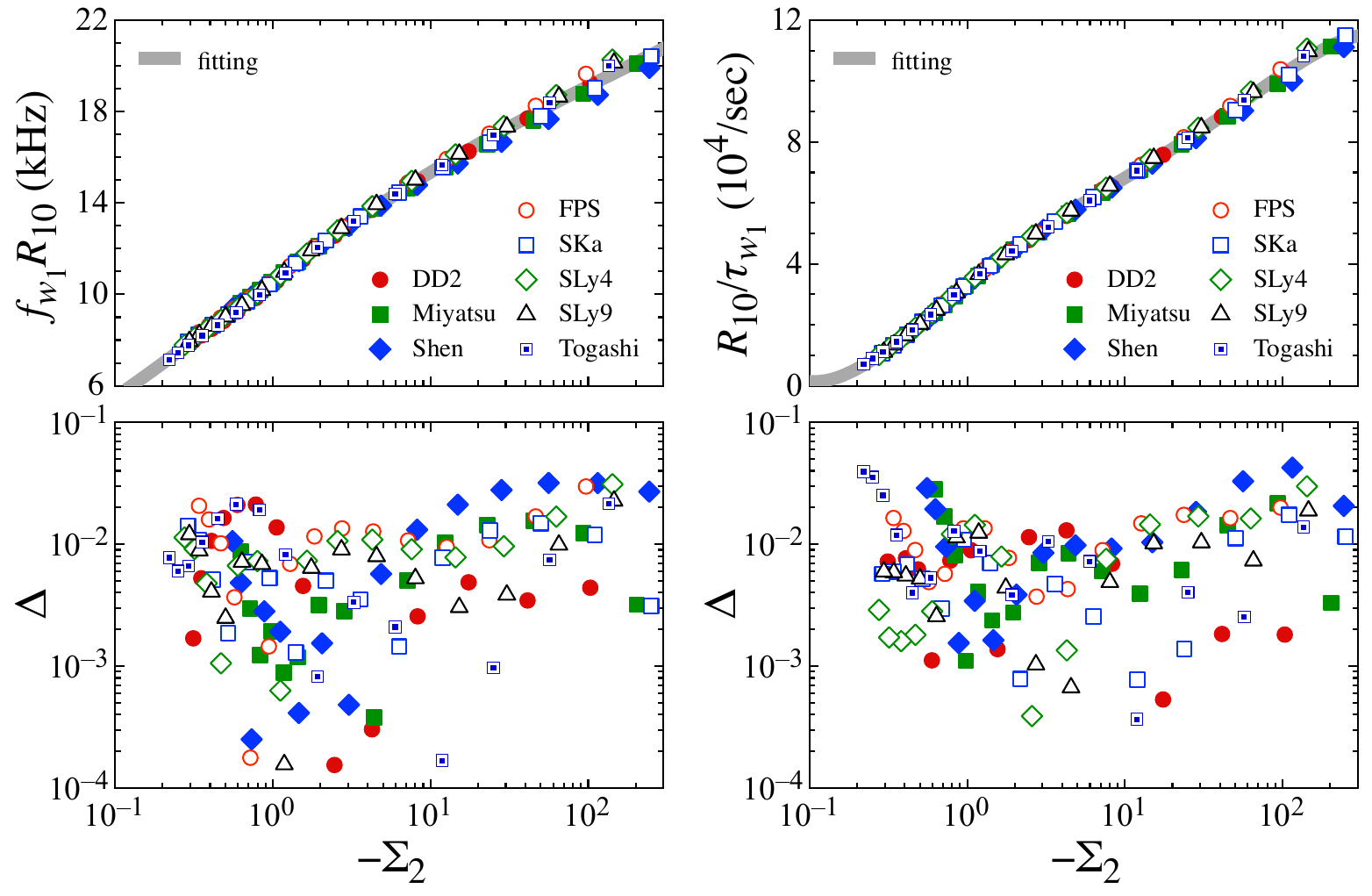} 
\end{center}
\caption{
Same as in Fig.~\ref{fig:ffM-ftauM}, but for the radius-scaled $w_1$-mode frequencies and damping rate with the fitting formula given by Eqs.~(\ref{eq:fw1M-Sigma}) and (\ref{eq:w1tauR-Sigma}). }
\label{fig:fw1R}
\end{figure*}

The $f$- and $p_1$-modes are gravitational waves strongly associated with the fluid motion, while the quasinormal modes primarily induced by the spacetime oscillations, also exist, i.e., the $w$-modes. In general, the $w$-modes are less sensitive to the fluid motion, but strongly depend on the stellar compactness, $M/R$. Just like the $p$-modes, the $w$-modes also exist infinitely. In this study, we focus on the 1st $w$-mode, i.e., $w_1$-mode, and examine its dependence on the magnetic tidal deformability. In Fig.~\ref{fig:fw1R}, we show the radius-scaled $w_1$-mode frequencies, $f_{w_1}R_{10}$ (top-left panel), and their radius-scaled damping rate, $R_{10}/\tau_{w_1}$ (top-right panel), as a function of $-\Sigma_2$, where $R_{10}$ is defined as $R_{10}\equiv R/(10\ {\rm km})$. In both panels, the solid lines denote the fitting formulae given by 
\begin{gather}
  f_{w_1}R_{10}\ ({\rm kHz})
   = 10.6033 + 5.1412x -0.2440 x^2 -0.2208 x^3+  0.060381 x^4, \label{eq:fw1M-Sigma} \\
  R_{10}/\tau_{w_1}\ (10^4/{\rm sec})
   = 3.3125 + 4.1740x -0.5427 x^2 -0.6683 x^3+ 0.7131 x^4-0.1665x^5. \label{eq:w1tauR-Sigma}
\end{gather}
The bottom panels show the absolute values of the relative deviation of the quantities shown in the above panels from the estimations with the fitting formulae. From this figure, one can estimate the $w_1$-mode frequencies and damping rate with an accuracy of at least $5\%$, using the fitting formulae.

In this study, we derive five universal relations associated with the quasinormal modes as a function of $-\Sigma_2$, i.e., Eqs.~(\ref{eq:ffM-Sigma}) -- (\ref{eq:w1tauR-Sigma}), which work well to estimate their frequencies and damping rate. Compared to the universal relations~\cite{SK21} as a function of the electric tidal deformability, $\Lambda_2$, the usefulness is more or less similar. In practice, the observation of the $w_1$-modes is more difficult than the $f$-modes, because the $w_1$-mode frequency is much higher and its damping rate is much shorter than those for the $f$-mode. Even so, using the empirical relations derived here, one may extract the imprints of the $w_1$-mode through the evaluation of $-\Sigma_2$ (and $\Lambda_2$), once the $f$-mode is detected.

At the end, we add a comment on the connection to observations of gravitational waves from neutron star mergers. Due to the tidal effect in the binary evolution, the gravitational waveform is modified by the electric tidal deformability, where a correction in the phase of the gravitational-wave signal with the frequency of $f_{\rm GW}$ is given by~\cite{Yagi14,PC21}
\begin{equation}
  \Psi_\ell = -\sum_{i=1}^2\left[\frac{5}{16}\frac{(2\ell - 1)!!(4\ell+3)(\ell+1)}{(4\ell-3)(2\ell-3)}\Lambda_{\ell}^{i}X_i^{2\ell-1}x^{2\ell-3/2}+\frac{9}{16}\delta_{\ell 2}\Lambda_2^i\frac{X_i^4}{\eta}x^{5/2} \right] +{\cal O}(x^{2\ell-1/2}), \label{eq:Psi1}
\end{equation}
where $i$ identifies the star among two stars in the binary system; $x\equiv (\pi M_{\rm T} f_{\rm GW})^{2/3}$ with the total mass $M_{\rm T}=M_1+M_2$; $X_{i}\equiv M_i/M_{\rm T}$; $\eta\equiv M_1M_2/M_{\rm T}^2$; and $\delta_{\ell\ell'}$ denotes the Kronecker symbol. On the other hand, the modification of the gravitational waveform due to the $\ell=2$ magnetic tidal deformability, which is a higher-order correction in the Post-Newtonian expansion rather than a correction due to electrical tidal deformation, is given by~\cite{Yagi14,PC21}
\begin{equation}
  \Psi_2 = \sum_{i=1}^2\frac{5}{224}\Sigma_2^i\frac{X_i^4}{\eta}(X_i-1037X_j)x^{7/2}+{\cal O}(x^{9/2}). \label{eq:Psi2}
\end{equation}
Separately from these finite-size effects of the neutron stars, the $f$-mode resonance with the orbital frequency also changes the gravitational waveform. So, to discuss the modification of the gravitational waveform with the $f$-mode frequencies and tidal effects, the universal relations between the $f$-mode frequencies and dimensionless tidal deformability may become useful.

\begin{figure}[tbp]
\begin{center}
\includegraphics[scale=0.4]{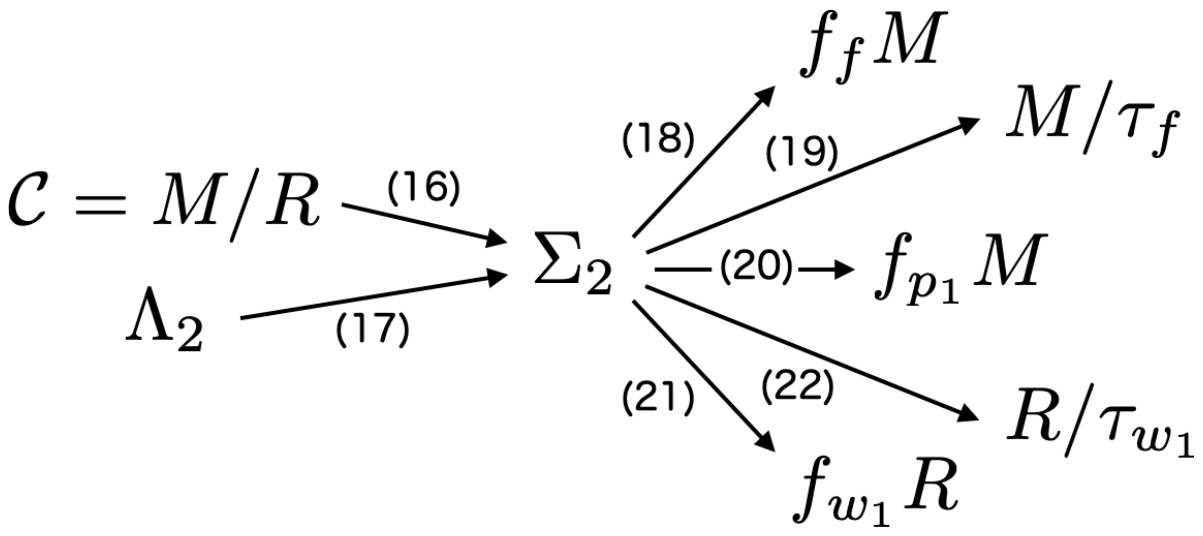} 
\end{center}
\caption{
Summary of the relation discussed in this study. The properties at the tip of the arrow are expressed as a function of the properties of the base of the arrow, where the number in brackets denotes the corresponding equation.}
\label{fig:relation}
\end{figure}

At the end, we summarize the universal relations discussed in this study. We focus on the electric and magnetic tidal deformability, $\Lambda_2$ and $\Sigma_2$; the stellar compactness, $M/R$; and quasinormal modes from the neutron stars. These properties would be constrained observationally as follows. The tidal deformabilities would be constrained from the gravitational waves from the inspiral phase, where $\Lambda_2$ has already been constrained from the GW170817, while $\Sigma_2$ might also be constrained in the future if the gravitational waves are precisely observed until the coalescence of binary neutron stars. The stellar compactness is primarily constrained from the pulsar light curve due to the relativistic light-bending effect. The quasinormal modes of a neutron star might potentially be detected in the future. Meanwhile,  the universal relations give us the theoretical connection between the various quantities. In practice, the relations discussed in this study are summarized in Fig.~\ref{fig:relation}. These relations enable us to estimate the other neutron star quantities using the observed ones, or to constrain the observed quantities more severely.

\section{Conclusion}
\label{sec:Conclusion}

Quasinormal modes excited in neutron stars are one of the most important observables to extract the properties of neutron stars in the context of gravitational wave asteroseismology. Direct observations of the quasinormal modes from an isolated cold neutron star may be quite challenging, but some of the modes, especially the $f$-mode, become important to understand the gravitational waveform from the neutron star binary mergers. In this study, we derive the universal relations expressing the $f$-mode frequencies and damping rate; the $p_1$-mode frequencies; and the $w_1$-mode frequencies and damping rate as a function of the magnetic tidal deformability. The universal relations derived in this study work well for estimating the frequencies and damping rate of the quasinormal modes, where the accuracy for the estimation is comparable to those as a function of the electric tidal deformability. In this study, we adopt the EOS for neutron star matter with relatively wide parameter ranges, but the universal relations could become more accurate if the EOS parameter range is constrained further in the future.

\acknowledgments
We are grateful to Ankit Kumar, Josuke Minamiguchi, Tomoya Uji, and Kenji Fukushima for their valuable comments. 
This work is supported in part by Japan Society for the Promotion of Science (JSPS) KAKENHI Grant Numbers 
JP23K20848         
and JP24KF0090. 





\end{document}